\documentclass[a4paper,11pt]{article}
\pdfoutput=1 % if your are submitting a pdflatex (i.e. if you have
\usepackage{jcappub}

\usepackage[T1]{fontenc} % if needed

\title{Prospects of Probing Dark Matter Condensates with Gravitational Waves}

\author[a,b]{Shreya Banerjee}
\author[c,d]{Sayantani Bera}
\author[e]{and David F. Mota}

\affiliation[a]{Inst. for Quantum Gravity, FAU Erlangen-Nuremberg,
Staudtstr. 7, 91058 Erlangen, Germany}
\affiliation[b]{Department of Physics, Ben-Gurion University, P.O.Box 653, Beer-Sheva 84105 Israel}
\affiliation[c]{Departament de Física, Universitat de les Illes Balears, IAC3 - IEEC, Crta. Valldemossa km 7.5, E-07122 Palma, Spain.}
\affiliation[d]{Inter-University Centre for Astronomy and Astrophysics, Post Bag 4, Ganeshkhind, Pune 411007, India.}
\affiliation[e]{Institute of Theoretical Astrophysics, University of Oslo, P.O. Box 1029 Blindern, N-0315
Oslo, Norway}

\emailAdd{shreya.banerjee@fau.de}
\emailAdd{sayantani.bera@uib.eu}
\emailAdd{d.f.mota@astro.uio.no}

\abstract{The Lambda-Cold Dark Matter model explains cosmological observations most accurately till date. However, it is still plagued with various shortcomings at galactic scales. Models of dark matter such as superfluid dark matter, Bose-Einstein Condensate(BEC) dark matter and fuzzy dark matter have been proposed to overcome some of these drawbacks. In this work, we probe these models using the current constraint on the gravitational wave (GW) propagation speed coming from the binary neutron star GW170817 detection by LIGO-Virgo detector network and use it to study the allowed parameter space for these three models for Advanced LIGO+Virgo, LISA, IPTA and SKA detection frequencies. The speed of GW has been shown to depend upon the refractive index of the medium, which in turn, depends on the dark matter model parameters through the density profile of the  galactic halo. We constrain the parameter space for these models using the bounds coming from GW speed measurement and the Milky Way radius bound. Our findings suggest that with Advanced LIGO-Virgo detector sensitivity, the three models considered here remain unconstrained. A meaningful constraint can only be obtained for detection frequencies $\leq 10^{-9}$ Hz, which falls in the detection range of radio telescopes such as IPTA and SKA. Considering this best possible case, we find that out of the three
condensate models, the fuzzy dark matter model is the most feasible scenario to be falsified/ validated in near
future. }


\begin{document}
\maketitle
\flushbottom

\section{Introduction}

The recent observational data in various Cosmological probes have substantiated the Lambda-Cold Dark Matter (LCDM) model and the success of the model has been firmly established with different independent  cosmological observations \cite{2009MNRAS.398.1150B,Iocco:2015xga,2014MNRAS.444.1518V}. For example, the model describes the Cosmic Microwave Background (CMB) temperature anisotropy, matter power spectra, large scale galaxy distributions and lensing observations with great accuracy \cite{Planck:2015bue}. Despite these achievements of this widely accepted model at cosmological scales, there are several inconsistencies at smaller scales that remain unresolved till date. Some of these issues are the so called cusp-core problem, Too Big To Fail problem, the Baryonic Tully-Fisher relation to mention a few \cite{2012MNRAS.422.1203B,2011MNRAS.415L..40B,2011ApJ...742...20W,Weinberg:2013aya,2012AJ....143...40M}.

The alternatives to LCDM come through modifications, either in the gravity sector or in the matter sector. The first class of models are called the modified gravity models like $f(R)$, $f(T)$, $f(\mathcal{G})$, Scalar-Tensor-Vector theories of gravity etc \cite{Starobinsky:1980te,Linder:2010py,Carroll:2003wy,Cognola:2006eg,Uzan:2010pm,Brownstein:2005zz,Horava:2009uw}. The other class of models arise due to a modification mostly to the matter content, especially by considering different kinds of Dark Matter (DM)  and Dark Energy (DE) \cite{2017AdSpR..60..166A}. In this paper, we focus on the modification coming from the second class of models which are modifications to the nature of Dark Matter . 

Various such models that concern a modification to the Dark Matter sector have been proposed earlier \cite{Bruneton:2008fk, Alexander:2018fjp,Ho:2010ca,Famaey:2017xou,Berezhiani:2015bqa}. Of primary interest are the interacting dark matter models such as the Bose Einstein Condensate(BEC) dark matter \cite{Suarez:2013iw} model and the superfluid dark matter model \cite{Berezhiani:2015bqa, Berezhiani:2015pia, Famaey:2017xou}. A very typical feature of these models is the formation of a BEC condensate at the core of the galaxies. Below a critical temperature, the DM particles undergo a phase transition at the galactic centre. Such models have been shown to successfully evade some of the issues posed by CDM at small scales. The forces are mediated by phonon modes and are described by an additional scalar field in the Lagrangian. In case of superfluid dark matter, originally proposed in \cite{Berezhiani:2015bqa}, it was shown to mediate a MOND-like interaction through phonon-baryon coupling. The superfluid represents a system with three-body interactions as opposed to usual BEC scenario where two-body interactions are mainly present. There is another kind of BEC DM proposed recently which are devoid of self-interactions \cite{Hu:2000ke,2020ApJ...889...88L}. They are commonly known as the Fuzzy Dark Matter models (FDM). Due to the ultralight mass of the DM particles in these models, they exhibit quantum phenomenon at the galactic core, and form solitons. These class of models have been widely tested at galactic and cosmological scales and prove to be promising DM candidates \cite{Hui:2016ltb}.\\
\indent  Gravitational wave sources have been used earlier in the context of possible merger of Proca stars which are self-gravitating boson stars comprising of fuzzy dark matter \cite{Carlos1,Carlos2,PhysRevD.99.044046}. Such compact dark matter objects have also been studied in \cite{PhysRevD.96.104058,Macedo_2013},\cite{Macedo_2013} focussing particularly on EMRIs (Extreme Mass Ratio Inspiral). Shapiro time delay has been proposed to probe the presence of dark matter substructures using pulsar timing observations \cite{10.1111/j.1365-2966.2007.12435.x,10.1093/mnras/stv2743,Khmelnitsky_2014,PhysRevD.98.102002}. The time delay observation coming from GW170817 has been used to rule out a class of modified gravity theories, called the \textit{dark matter emulators} \cite{Boran:2017rdn}. The LIGO-Virgo scientific collaboration has also constrained a certain class of axion dark matter termed as the dark photon dark matter using data from the third observing run \cite{LIGOScientific:2021ffg}. The primordial black holes(PBH) are also studied as viable candidate for dark matter and the detection of GW signal emitted from such sources has been considered earlier \cite{PhysRevD.92.023524,Raidal:2017mfl,PhysRevLett.119.131301,PhysRevLett.122.211301,PhysRevLett.122.041103}. In a different context concerning macroscopic gravity, the GW propagation dealing with modified dispersion relation and extra degrees of polarization in extended theories of gravity has been looked into in \cite{Montani:2018iqd,Moretti:2020kpp,Moretti:2021ljj,Moretti:2022xem}. In this paper, we rather focus explicitly on DM condensate candidates that serve as constituent candidates for galactic DM halo. We examine the prospect of constraining the various model parameters of the DM models mentioned in the paragraph above using the gravitational wave (GW) speed measurements coming from the LIGO-Virgo observations \cite{Cai,Dev:2016hxv,LIGOScientific:2018mvr,TheLIGOScientific:2017qsa}. We specifically use the constraints on the relative change of speed of GW from the observation of the binary neutron star (BNS) merger GW170817 \cite{LIGOScientific:2018mvr,TheLIGOScientific:2017qsa} to probe the allowed parameter space of these models. When the GW passes through a medium different from vacuum, its interaction with the medium gives rise to a phase shift, thereby resulting in an effective change in the speed in this medium. This has been discussed in detail in \cite{Peters:1974gj}. We use this idea to study the effect of these BEC media on the GW as it passes through them. Further, the electromagnetic waves while passing through such media do not undergo any change in speed with respect to vacuum, as noted in \cite{Peters:1974gj}. For an overview on the subject of how gravitational wave observation can be used as a probe for different classes of dark matter, one can, for example, look up \cite{Bertone:2019irm, Barack_2019,PhysRevResearch.1.033187}. \\
\indent The paper is organized as follows: In section \ref{sec2}, we describe the general idea of this work in brief. Section \ref{sec3} discusses the analysis for the case of superfluid DM, both in presence and absence of baryonic interactions. The results for the FDM and the BEC DM have been investigated and discussed in sections \ref{sec4} and \ref{sec5} respectively. Finally, in section \ref{sec6}, we summarize our main results.

\section{Probing Dark Matter models through Gravitational Waves}
\label{sec2}

 When a gravitational wave passes through a medium, it can interact with the matter and the resulting outgoing wave can undergo a phase shift depending upon the matter density of the medium. This is similar to the case of electromagnetic waves passing through a medium where the light-matter interaction causes a change in effective refractive index of the medium. In case of tensor fields like gravitational waves, there have been a number of efforts to capture this behaviour through various kinds of treatment of the effect of the gravitational wave on the medium of propagation and its backreaction on the waves itself \cite{SZEKERES1971599,PhysRevD.7.2863,Peters:1974gj}. Most of these approaches deal with the weak field regime of gravity, since working with linearized gravity immensely simplifies the calculations by enabling the addition of fields generated by single particles. In the majority of the cases of interest, the weak field approximation works well. In this paper, we are concerned with the approach followed in \cite{Peters:1974gj}, where the propagation of scalar, vector (electromagnetic waves) and tensor fields (gravitational waves) through a gravitating medium and its effect on each kind of field has been studied and an attempt to draw an analogy between the three cases have been made. For the gravitational waves, it has been shown in \cite{Peters:1974gj}, that the change in effective refractive index is proportional to the average density of the medium and is also dependent upon the frequency of the gravitational wave.\\

The description is given in the weak field regime of gravity. In the weak field limit, the metric perturbations to a flat Minkowskian spacetime can be written as,
\begin{eqnarray}
    g_{00} = 1+2\phi ; \nonumber \\
    g_{0i} = 0; \nonumber \\
    g_{ij} = -\delta_{ij}(1-2\phi)
    \label{metric}
\end{eqnarray}
where $\phi$ is the Newtonian potential. For a particle of mass $m$, this is simply $\phi= -Gm/r$. The potential for a mass distribution can be obtained by summing over the potential due to each particle, since we are in the linearized gravity regime.\\

In the weak field limit, the perturbed background metric in the presence of gravitational waves can be written (up to first order) as :
\begin{equation}
    g_{\mu\nu} = g_{\mu\nu}^{(0)} +  h_{\mu\nu} 
\end{equation}
where $g_{\mu\nu}^{(0)}$ is the metric described in Eq. \eqref{metric}.\\
One can derive the wave equation starting from this perturbed metric \cite{Peters:1974gj}. For our case of interest, we rather focus only on the spatial components of the full wave equation. Using the first-order perturbation approximation and assuming that the polarization vector is constant, one arrives at the following equation for the amplitude $h$ of the gravitational waves \cite{Peters:1974gj}.
\begin{equation}
    (\nabla^2 + \omega^2)h = 4\omega^2\phi h
    \label{gweq}
\end{equation}
where we have used a co-ordinate dependence for $h$ of the form $h(\textbf{r},t)=h(\textbf{r}) e^{-i\omega t} $.\\

The right hand side of Eq. \eqref{gweq} represents the interaction of the GW with the dark matter medium. For a single particle, the scenario is equivalent to the scattering of the GW off the particle, and the resultant wave would be a superposition of incident and scattered wave, and as a consequence, the phase of the outgoing wave would be modulated. For a system with many particles, one needs to consider multiple scattering processes within the medium. This exercise is non-trivial, but in the weak gravity regime, since the potential can be written a linear sum of potentials due to each particle, one can just add up the scattering effects arising due to each particle. Consider an incident GW, $h \sim e^{i kz}$ propagating in the $z$-direction and encountering a scattering medium at $z=0$. Then for $z>0$, the total wave can be written as $$h_{tot} \sim e^{i(kz+\delta\Phi)}$$ where $\delta\Phi$ is the phase change due to the interaction of the GW with the scattering medium. If the thickness of the medium is $\Delta z$, then this phase change can be obtained by calculating how much time the GW takes to traverse the distance $\Delta z$ in the presence of the medium in comparison to vacuum. The phase velocity inside a medium of refractive index $n$ is given by $c_g = 1/n$, taking the speed of light in vacuum as $c=1$. So, the phase difference due to a medium of thickness $\Delta z$ would be,
\begin{equation}
    \delta \Phi = \omega \Delta z n - \omega \Delta z 
\end{equation}
If one considers $\Delta z$ to be very small, the phase change would be small enough so that we can use the approximate expansion of the form $$ e^{i \delta \Phi} \approx 1 + i\delta \Phi $$
Under this approximation, the total outgoing wave in the region $z>0$ can be written as,
\begin{equation}
    h_{tot}  \sim e^{i kz} [1+i(n-1)\omega \Delta z]
    \label{gw1}
\end{equation}

In principle, there will be another contribution to the outgoing wave arising due to focusing effects as a result of deflection, as discussed in \cite{Peters:1974gj}. But that effect is not related to the effective change of speed of the wave inside the medium, so we do not concern ourselves with the contribution. The main interest for our case is to track any changes to the waveform as a result of the effective change of speed. The solution for the GW wave equation Eq. \eqref{gweq} can be found by setting $\omega = k$ and keeping terms only up to order $\phi$. The solution is a hypergeometric function, which can be further simplified by expanding in order of $G \rho$ and keeping terms up to first order. There is a logarithmic phase factor in the solution  added with respect to the incoming wave. If one imposes the condition, that in high frequency-limit, the phase velocity should be same as the speed of light, then the logarithmic factor is eliminated, and the solution becomes (without focusing term) :
\begin{equation}
    h \sim e^{ikz} \left(1+\frac{2 \pi G \rho i\Delta z}{\omega}\right)
    \label{gw2}
\end{equation}
    
Comparing Eq. \eqref{gw2} with Eq. \eqref{gw1}, one finds that
the refractive index of the medium is related to the density by the following expression-
\begin{equation}
    n=1+\frac{2\pi G \rho}{\omega^2}
    \label{5}
\end{equation}
where $\omega$ is the frequency of the incident GW wave and $\rho $ is the density of the medium.\\

For a density that varies, one can take the average density $\langle \rho \rangle$ of the medium for any arbitrary thickness $z$ such that $$\langle \rho \rangle = \frac{1}{z}\int_0^z \rho(z') \,dz'$$ The average density would depend upon the density profile, and has to be computed for each kind of density profiles separately. More on this calculation has been discussed in the next section.

As a consequence of the change in refractive index as seen by the GW, the resultant wave captured at the detectors can have a speed that is significantly different from the speed of light in vacuum. If we consider gravitational waves emanating from a distant source at a distance $D$, then while passing through a dark matter halo, the average fraction of distance inside the halo would be $$x = \frac{\langle D_h \rangle}{D}$$ where $\langle D_h \rangle$ is the average of all the distances inside a halo (for a spherical halo of radius $R$, $D_h$ ranges from 0 to $2R$). For a spherical halo of radius $R$, this fraction is $\sim R/D$. Since this fraction of the total distance is traversed with a reduced speed $c_g = 1/n$, the total time can be found as $$\Delta T = \frac{R}{c_g} + (D-R)$$ So, once can define an effective speed as the total distance divided by the time $\Delta T$, $c_{eff}$ (for $c=1$):
\begin{equation}
    c_{eff} = \frac{D}{\frac{R}{c_g} + (D-R)}
    \label{speed1}
\end{equation}

The observation of the gravitational waves from the binary neutron star GW170817 detected in the LIGO-Virgo detectors along with its electromagnetic counterpart already constrains the relative change of speed of the GW with respect to light to a high degree of accuracy which has been reported to be $\sim 10^{-15}$ \cite{Monitor:2017mdv}. 

In this work, we consider this change of speed in the presence of interacting dark matter in our Milky Way galaxy. In particular, we study three different models, namely, a) superfluid dark matter, b) fuzzy dark matter and c) BEC dark matter to constrain their respective model parameters.

\section{Superfluid Dark Matter}
\label{sec3}

The model of superfluid Dark Matter was proposed by Khoury and Berezhiani in \cite{Berezhiani:2015bqa} in order to address the galactic scale issues of the usual CDM. The main idea of the model is that the dark matter particles can form a condensate and undergo phase transition below a critical temperature determined by the scale under consideration. Below this temperature, the condensate acts as a superfluid medium with phonon modes mediating the MOND-like interactions through coupling with baryons. 

The total Lagrangian for this interacting superfluid DM model in the low energy effective theory is given as,
\begin{equation}
{\cal L} = \frac{2}{3} \Lambda(2m)^{3/2}X\sqrt{|X|} - \alpha\frac{\Lambda}{M_{\rm Pl}} \theta \rho_{\rm b}\,,
\label{1}
\end{equation}
where $X =  \dot{\theta} - m\Phi - (\vec{\nabla}\theta)^2/2m$ , and $\rho_{\rm b}$ is the baryon density.\\

 Here, $\theta$ represents the phase of the wavefunction describing the phonon modes, $\Phi$ is the standard Newtonian gravitational potential in the usual non-relativistic case. The free parameter $m$ is the mass of the DM particle, $\Lambda$ describes the range of the superfluid, and the other free parameter $\alpha$ represents the strength of baryon-superfluid interaction.It has been suggested in \cite{Berezhiani:2015bqa} that the bounds on the model parameters should be different at different length scales, thus implying their different behaviour at large scales(pressure-less fluid) and small scales (MONDian force mediating superfluid phonons). Bounds on $\Lambda$ and $\alpha$, for a given $m$ have been obtained at the galactic and cosmological scales in \cite{Berezhiani:2015bqa, Berezhiani:2015pia, Banerjee:2020obz}.
 The fractional power of the Lagrangian is somewhat arbitrary but motivated by the choice of the equation of state (EoS) and the fact that the superfluid DM should give rise to MOND-like dynamics at galactic scales. The corresponding equation of state is 
 \begin{equation}
   P = \frac{\rho^3}{12 \Lambda^2 m^6}  
   \label{eos}
 \end{equation} 
 The dependence of the pressure $P$ arising due to self-interaction on the density $\rho$ suggests that the nature of the interaction between the particles is predominantly a three-body interaction process, different from a BEC with an equation of state $P \sim \rho^2$ \cite{Berezhiani:2015bqa}. This means that the potential term representing the interaction cannot be expressed as a sum of pairwise interactions, rather a three-body interaction between the particles has to be taken into account in order to describe the dynamics of the system.\\
 
 In the following section, we consider two different cases for the superfluid DM, a) without baryonic interactions and b) with baryonic interactions to study the relative change of GW speed when it passes through the superfluid medium.
\subsection{Without Baryonic interactions}

In the absence of baryonic interactions, the Lagrangian takes the form,
\begin{equation}
{\cal L} = \frac{2}{3} \Lambda(2m)^{3/2}X\sqrt{|X|}
\label{2}
\end{equation}

The density profile can be obtained by solving the pressure equation for a static, spherically symmetric halo, given by,
\begin{equation}
    \frac{1}{\rho(r)}\frac{dP(r)}{dr} = -\frac{-4\pi G}{r^2}\int_0^r r'^2\rho(r')dr'
\end{equation}

This can be expressed in the form of the Lane-Emden equation \cite{Berezhiani:2015pia} as
\begin{equation}
\left(\xi^2 \Xi'\right)' = - \xi^2 \Xi^{1/2}
\label{3}
\end{equation}
by substituting $\rho$ and $r$ with the dimensionless variables $\Xi$ and $\xi$ respectively, defined as, $\rho = \rho_0\Xi^{1/2}$ and $r = \sqrt{\frac{\rho_0}{32\pi G \Lambda^2 m^6}}\,\xi$, with $\rho_0 $ denoting the central density of the DM halo.\\

Eq. \ref{3} is solved for the boundary conditions $\Xi(0) = 1$ and $\Xi'(0) = 0$. Numerical solution gives the density profile of the DM condensate halo. The solution vanishes at $\xi_1 \simeq 2.75$, which defines the halo size as $R = \sqrt{\frac{\rho_0}{32\pi G \Lambda^2 m^6}}\; \xi_1$.
The central density is related to the halo mass, $M$ as $\rho_0 =  \frac{3M}{4\pi R^3} \frac{\xi_1}{|\Xi'(\xi_1)|}$,
with $\Xi'(\xi_1)\simeq -0.5$. From \cite{Berezhiani:2015pia}, we get the following expressions for $\rho_0,\ R$:
\begin{eqnarray}
\nonumber
\rho_0 &\simeq & M_{12}^{2/5} m_{\rm eV}^{18/5}\Lambda_{\rm meV}^{6/5} \; 7 \times 10^{-25}~{\rm g}/{\rm cm}^3  \,; \\
R &\simeq & M_{12}^{1/5} m_{\rm eV}^{-6/5}\Lambda_{\rm meV}^{-2/5} \; 36~{\rm kpc}\,,
\label{4old}
\end{eqnarray}

where $M_{12} = M/10^{12} M_{\odot}$, $m_{\rm eV}$ is the DM particle mass in eV units and $\Lambda_{\rm meV}$ is the model constant $\Lambda$ expressed in meV units. 

The above relations give the central density and the size of the superfluid core in the DM halo. In practice, the halo is expected to consist of a superfluid core surrounded by ordinary CDM crust following a NFW (Navarro-Frenk-White) profile. The general density profile of the DM halo would have the following form:
\begin{equation}
 \rho(r) =\rho_{c}(r)\theta(r_{t}-r)+\rho_{NFW}(r)\theta(r-r_t),
 \label{fuzzyold}
 \end{equation}
where $\theta$ is a step function and $r_t$ is  the scale/radius at which the transition from superfluid to ordinary CDM phase happens. The first term corresponds to the core density of the superfluid dark matter halo profile up to radius $r_t$ whereas the second term gives the density profile between $r_t$ and the halo radius $R$.
Here
$\rho_{NFW}$ corresponds to the NFW density profile \cite{NFW} in the halo region which is given as
\begin{equation}
    \rho_{NFW}(r)=\frac{\rho_{NFW}^0}{r/R_{NFW}\Big(1+(r/R_{NFW})^2\Big)}
    \label{nfweq}
\end{equation}

For such a generic halo profile with central superfluid core surrounded by a CDM crust, the core central density $\rho_0$ and core radius $R_{\rm core}$ can be computed using \eqref{4old} where the mass $M_{12}$ will be replaced by the core mass.

The relation between the refractive index and the density of the medium for passing GW wave is given in Eq. \eqref{5}. 
The equation is valid when the density $\rho$ is constant. When we have a medium for which the density profile evolves as a function of the distance, for a very thin layer (of width $r$) of the medium such that the density $\rho$ can be taken to be constant within that layer, Eq. \eqref{5} can be written as :
\begin{equation}
    n(r)=1+\frac{2\pi G \rho(r)}{w^2}
    \label{5a}
\end{equation}
In this case, the refractive index changes from one layer to another. Thus, when we have a medium of width $R$ and density at any distance $r$ is given as $\rho(r)$, then the effective refractive index of the medium as seen by the incident GW with frequency $w$ would be given as follows,

\begin{equation}
    n=1+\frac{2\pi G \int_0^R\rho(r) dr}{w^2 \int_0^R dr}
    \label{5b}
\end{equation}
The second term in the R.H.S denotes the modification of the refractive index with respect to vacuum. This is equivalent to using $\langle \rho \rangle$ discussed previously, instead of $\rho$ in Eq. \eqref{5}. \footnote{Note: Eq. \eqref{5} has been derived in \cite{Peters:1974gj} for a medium with thickness $\Delta z$, $\Delta z$ being very small. Thus, the formula is valid only for very thin shells. So, we split the medium of a certain thickness $z$ into very thin shells of thickness $\Delta z$, so that for each shell, the formula holds true. Combining all such shells and taking the limit $\Delta z \to 0$, we get the integration of the form given in Eq. \eqref{5b}. The result can be thought of as considering an average density $\langle \rho \rangle$ for the halo, instead of $\rho$ that varies with positions.}\\

We assume that the dominant contribution to the modification comes due to the DM halo of the Milky way galaxy, and the rest is just vacuum on an average (this is an oversimplification of a more complicated problem, but we expect that the order of magnitude would not change much otherwise).\\

The superfluid core density profile can be expressed in terms of dimensionless variables $\Xi$ and $\xi$ as described above. In terms of the new dimensionless variables $\Xi$ and $\xi$  we can write the core density profile as,
 \begin{equation}
     \rho_c = \rho_0 \Xi^{1/2}
 \end{equation}
 and \begin{equation}
     r = C_1 \xi
 \end{equation}
 where $C_1=\sqrt{\frac{\rho_0}{32\pi G \Lambda^2 m^6}} $. \\
 
 Substituting the above results in Eq. \eqref{5b} and using the full halo profile as given in \eqref{fuzzyold}, we get,
 
 %\vspace{0.2in}
 \begin{eqnarray}
     n = 1+\frac{2\pi G}{w^2 R}\left[\int_0^{r_t}\rho_c(r) dr +\rho_{NFW}^0\int_{r_t}^{R}dr\frac{1}{r/R_{NFW}\Big(1+(r/R_{NFW})^2\Big)}\right]
     %\label{ri}
 \end{eqnarray}
Expressing the superfluid density $\rho_c$ in terms of dimensionless variables $\Xi$ and $\xi$, we have,
\begin{equation}
     n = 1+\frac{2\pi G}{w^2 R}\left[\rho_0 C_1 \int_0^{r_t/C_1}\Xi^{1/2} d\xi +\rho_{NFW}^0\int_{r_t}^{R}dr\frac{1}{r/R_{NFW}\Big(1+(r/R_{NFW})^2\Big)}\right]
     \label{ri}
 \end{equation}
If $D$ be the total distance that the GW travels i.e. the distance between the observer (us) in the Milky way and the merger (considering the medium in between is vacuum), then the average fraction of distance the GW propagates through the halo with a reduced speed is $R/D$. The effective speed of the GW can then be written from Eq. \eqref{speed1} as
\begin{equation}
    c_{eff}=\frac{c_g}{\frac{R}{D}+(1-\frac{R}{D})c_g}
\end{equation}
where $c_g$ is the speed of the gravitational wave through a medium. \\

Therefore, the change in speed of GW compared to the speed of light in vacuum (assuming $c=1$) is given as
\begin{equation}
\delta c_{eff}= \frac{\delta c_{g}}{\frac{R}{D}+(1-\frac{R}{D})c_g}-\frac{c_g(1-\frac{R}{D})\delta c_g}{(\frac{R}{D}+(1-\frac{R}{D})c_g)^2}
\label{ceff}
\end{equation}

We also know that the reduced speed of the GW is inversely proportional to the refractive index of the medium ($c_g \sim 1/n$). \\
 Neglecting higher order terms in $\delta n$, 
 $$\delta c_g \approx -\delta n$$
 
This relation is used while computing $c_{eff}$.\\
 
{\bf Observational data from GW170817}:\\

We compute the effective speed $c_{eff}$ for typical values obtained in case of the BNS source GW170817. Below we state the observed values for the various parameters of the source \cite{TheLIGOScientific:2017qsa,LIGOScientific:2018mvr}.\\

 $$w \sim 100 ~~ Hz \footnote{ This is an order of magnitude proxy for the signal frequency which ranges roughly between 50 Hz and 500 Hz}$$ 
 $$D_L = 40 ~~Mpc \quad \quad z= 0.01$$
 Thus, the physical distance to the source $D= \frac{D_L}{1+z}$ is  $39.6 \times 10^3$ kpc.\\
 
For the NFW profile, we use the following parameters for a Milky Way like galaxies found in the literature \cite{NFW}:\\

$\rho_{NFW}^0=2.9\times 10^{-2}\ M_{\odot}/pc^3; \quad R_{NFW}=10^4 pc; \quad R=10^5 pc$ \\

 The strongest bound on the relative change in speed of GW has been observed as $\delta c_{eff} \sim 10^{-15} $ with $c=1$ \cite{Monitor:2017mdv}. \\
 
\begin{figure}[h!]
\centering
\includegraphics[width=4in]{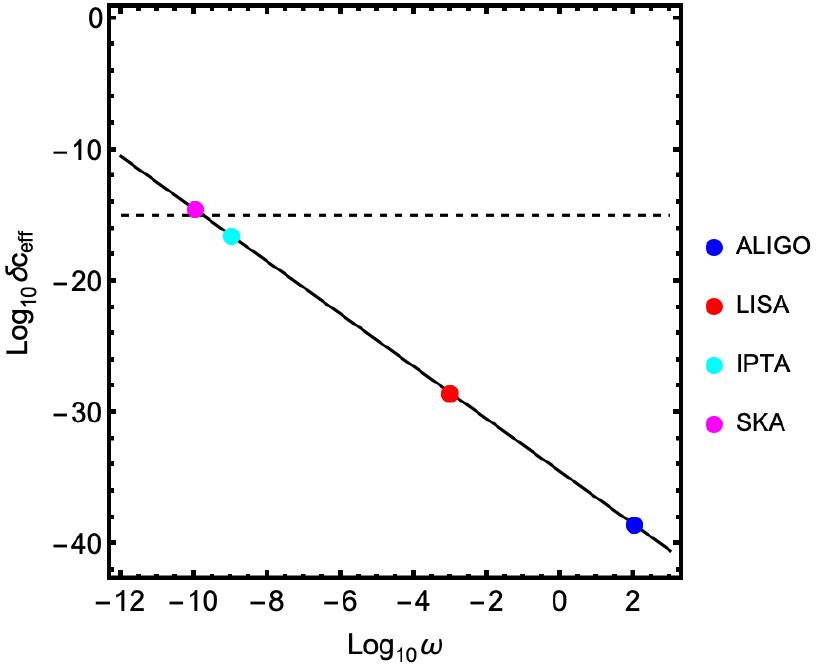}
\caption{{\it Variation of change in the effective speed of GW ($\delta c_{eff}$) with frequency (w), given by the solid line, for superfluid dark matter without baryonic interaction with $\Lambda=10^3$ meV and $m=5\times 10^{-2}$ eV. We assumed that the ratio of superfluid core mass to total halo mass is $0.9$. The dotted line corresponds to the current upper limit on $\delta c_{eff} \sim 10^{-15}$ put by LVK from the multi-wavelength detection of GW170817 in EM and GW frequencies. The coloured dots represent the change in effective GW speed for ALIGO, LISA, IPTA and SKA detection frequencies.} }
\label{plot1}
\end{figure}

 For the BNS source GW170817, we use the data as above, and for the merger frequency $w = 100$ Hz, the relative change in the effective speed of the GW is calculated to be, using \eqref{ceff}, $\delta c_{eff} \sim 10^{-39} $ for the current (advanced) sensitivity of LIGO-Virgo detectors (referred to as ALIGO in short hereinafter). This estimate is much smaller than the precision of the current detectors, smaller by almost 25 orders of magnitude. Detection of GW at a lower frequency would improve the estimate since the change in refractive index is inversely proportional to the square of the detection frequency. For sources detected by space-borne telescope LISA, the highest sensitivity is achieved at a frequency of $w \sim 10^{-3}$ Hz, for which the fractional change in speed is estimated to be $\sim 10^{-29}$. While this is a much larger effect in comparison to ALIGO, the order-of-magnitude is still 15 orders less than the detectable range. A detectable change in the speed can be achieved at a much lower frequency $\sim 10^{-9}$ Hz, which is typical range for radio telescopes such as International Pulsar Timing Array (IPTA) \cite{Antoniadis:2022pcn,2016MNRAS.458.1267V,2019MNRAS.490.4666P,Kelley:2017lek} and Square Kilometer Array (SKA) \cite{Janssen:2014dka,Moore:2014lga}, for which the corresponding change in speed is of the order $10^{-15}$. The effect of detection frequency on the relative change in speed is shown in Fig. \ref{plot1}. The dotted line corresponds to the current upper limit on $\delta c_{eff}$ put by LVK from the multi-wavelength detection of GW170817 in EM and GW frequencies. The solid black line shows the effective change in speed with frequency for the superfluid dark matter without baryonic interactions, assuming the superfluid core mass to be $90\%$ of the total halo mass, and fixing the DM parameters at $\Lambda=10^3$ meV and $m=5\times 10^{-2}$ eV. The coloured dots represent the change in effective GW speed for ALIGO, LISA, IPTA and SKA detection frequencies. As can be seen in the plot, only SKA is above the detectable range while LISA and ALIGO are much below the threshold. \\
 
\begin{figure}[h!]
\centering
\includegraphics[width=5in]{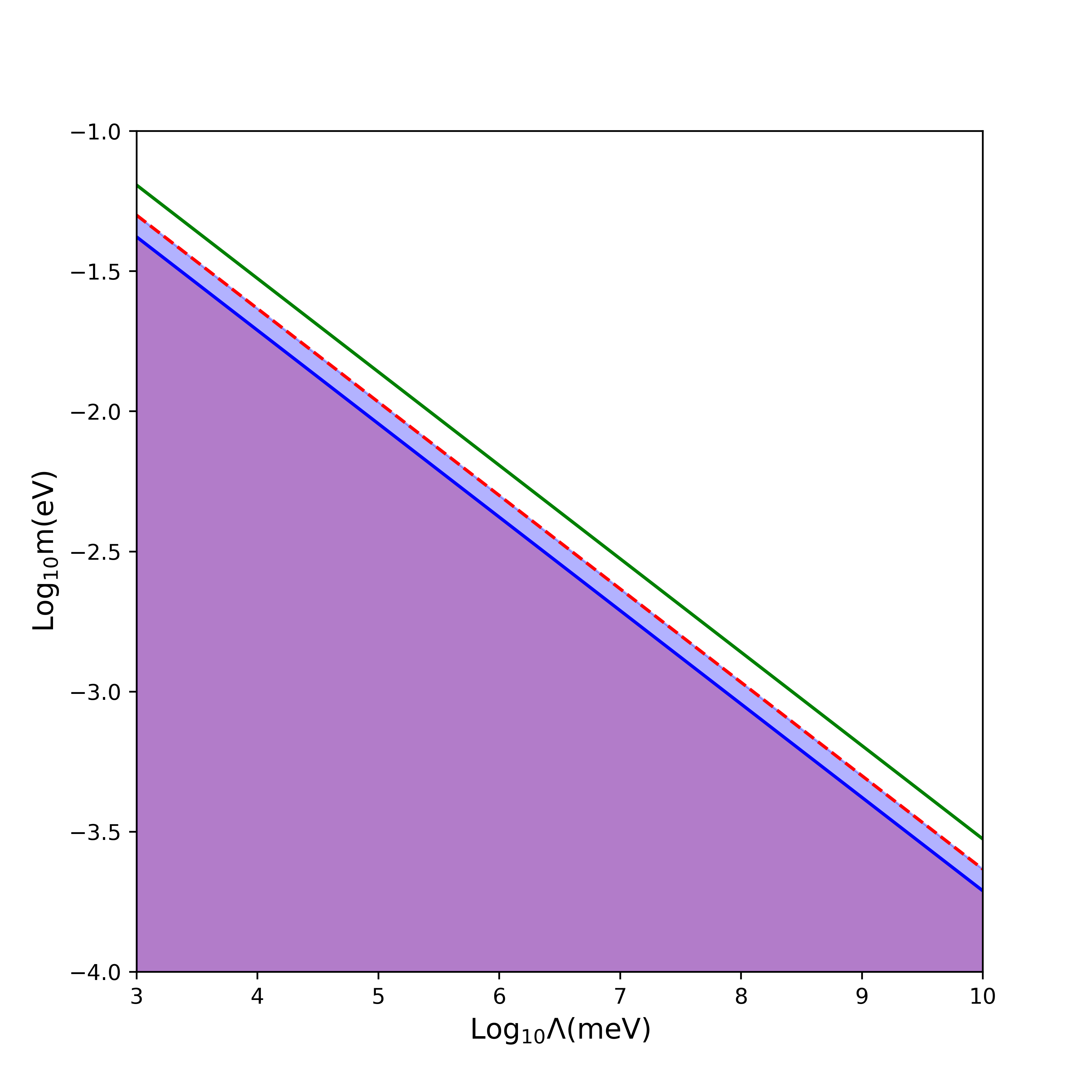}
\caption{{\it Allowed parameter space for superfluid dark matter without baryonic interaction with frequency fixed in the nanoHz range. The shaded region represent the parameter space allowed by the current constraint on the relative change in speed. The red dashed line corresponds to the upper limit of the allowed parameters coming from the GW speed constraint. The blue line represents the parameter values defined by $R=100$kpc and the green line is for $R=60$kpc. The blue shaded region is the allowed region that satisfies both speed and radius constraints.} }
\label{plot11}
\end{figure}

 We also plot the allowed parameter space for the superfluid dark matter considering the best case scenario where the frequency is fixed in the nanoHz range. For this case, the density profile transits from superfluid core density profile to an NFW profile smoothly, and the transition radius is assumed to be at $r_t \sim 60$ kpc. This is motivated by the fact that for a Milky Way like galaxy, the rotation curves are observed out to $60 $ kpc and this observation must be satisfied by the superfluid DM. This requires the transition radius to be at least $60$ kpc, as noted in \cite{Berezhiani:2017tth}. Requiring the fact that the superfluid core thus has to be at least $60$ kpc, we get an upper bound in the parameter space. In Fig. \ref{plot11}, we show the parameter space allowed /ruled out by the relative change in speed. The shaded represent the allowed region by the GW speed measurement.
 From \eqref{4old}, for a given radius, we get a relation between $\Lambda$ and $m$. Requiring that the core radius has to be less than the Milky Way radius ($ \sim 100$ kpc), but greater than the transition radius $60$ kpc, we have an additional constraint on the model parameters. The blue line indicates the parameter values defined by $R=100$ kpc and the green line corresponds to $R=60$ kpc. The red dashed line corresponds to the upper limit on $\delta c_{eff}\sim 10^{-15}$ imposed by the GW speed constraint.  The blue shaded area bounded by the red and blue lines represent the area allowed by both speed and radius constraints. We thus see that there are a number of parameter values that are allowed by the radius constraint but are disfavoured by the gravitational wave speed bound.


\subsection{With Baryonic interactions}

In this section, we repeat the analysis, in the presence of the interaction term in the superfluid Lagrangian \eqref{1}. In this case, we have an additional contribution to the Lane-Emden equation coming from the baryon-phonon interaction. 
The Lane Emden equation, in this case, is given as,
\begin{equation}
\frac{1}{\xi^2}\frac{\mathrm{d}}{\mathrm{d}\xi}\left(\xi^2\frac{\mathrm{d}\Xi}{\mathrm{d}\xi}\right)=-\Xi^\frac12-\frac{\rho_0^\frac12}{\alpha m\Lambda}\frac{1}{\xi^2}\frac{\mathrm{d}}{\mathrm{d}\xi}\left(\xi^2\Xi\right).
\label{int}
\end{equation}

Here in this case, using the Lane-Emdane equation above, we find the central density $\rho_0$ of the core as follows,
$$\rho_0\int \Xi^{1/2} dV = M$$
Using $dV = r^2 d\Omega dr = C_1^3 \xi^2 d\Omega d\xi $, and substituting this in the integral above, one gets,
$$\rho_0C_1^3 \int \Xi^{1/2} \xi^2 d\xi = \frac{M}{4\pi} $$
From Eq. \eqref{int}, we thus get,
$$\rho_0 C_1^3 \left(\xi_1^2 |\Xi'(\xi_1)| + \frac{\rho_0^{1/2}}{\alpha m \Lambda}\xi_1^2 |\Xi(\xi_1)|\right) = \frac{M}{4\pi}$$
The second term in the above expression goes to zero since $\Xi(\xi_1) = 0$. We thus end up with the following expressions for $\rho_0$ and the core size $R_{\rm core}$:
\begin{eqnarray}
\nonumber
\rho_0 &\simeq & M_{12}^{2/5} m_{\rm eV}^{18/5}\Lambda_{\rm meV}^{6/5} \; 2.64 \times 10^{-24}/(\xi_1^2|\Xi'(\xi_1)|)~{\rm g}/{\rm cm}^3  \,; \\
R_{\rm core} &\simeq & M_{12}^{1/5} m_{\rm eV}^{-6/5}\Lambda_{\rm meV}^{-2/5} \; 25.45/|\Xi'(\xi_1)|^{1/2}~{\rm kpc}\,,
\label{4}
\end{eqnarray}
which have no direct dependence on the model parameter $\alpha$.

Since we do not apriori know the value of $\rho_0$ without knowing the solution for $\Xi$, Eq. \eqref{int} can only be solved iteratively. We start with a guess value for $\xi_1$ and insert it in Eq. \eqref{int} and check if the new value of $\xi$ obtained by solving the equation is close (within a precision) to the original value that we started with. We keep repeating this process until we reach a converging solution. A natural choice for the guess value would be the solution without the second term in Eq. \eqref{int} i.e. without the interaction term. The method is equivalent to the one adopted in \cite{Cai} and is described in detail below.\\

We solve Eq. \eqref{int} without the second term in the R.H.S, i.e, for the case when there is no interaction. The solution gives $\xi_1$, the value at which $\Xi(\xi) =0$. Using this $0^{th}$ order solution, we obtain the value of $\rho_0$ for a given set of parameter values $\Lambda$, $\alpha$ and $m$. With this $\rho_0$ now, we again solve for $\Xi$ using Eq. \eqref{int}. The new solutions are again plugged in to the next iteration where $\rho_0$ is again modified. In this way, at different iterations we solve for $\Xi$ until the value of $\xi_1$ converges. If the convergence is achieved at the $n^{th}$ step, then the final value of the halo central density and halo radius would be $\rho_0^{(n)}$ and $R^{(n)}$ respectively. Using this final solution, we now compute the relative change in speed of GW as discussed in the previous section. However, we find that the outcome is similar to the case without the presence of baryonic interactions i.e. the inclusion of baryonic interactions, hence the parameter $\alpha$, does not change the outcome significantly. We therefore conclude that our results are independent of the interaction parameter $\alpha$ as we do not get any new additional constraints on $\alpha$ from this analysis.

\section{Fuzzy Dark Matter}
\label{sec4}

The fuzzy dark matter (FDM) model has been proposed as a possible resolution for the core-cusp problem in galaxies. This model assumes the existence of ultra-light bosonic particles ($m \sim 10^{-23}-10^{-21}$ eV \cite{PhysRevD.28.1243,PhysRevLett.64.1084,1994PhRvD..50.3650S,2000PhRvL..85.1158H,Goodman:2000tg,Peebles_2000,2006PhLB..642..192A,2014NatPh..10..496S}) with no self-interaction. The particles form a BEC which has a large De-Broglie wavelength.  They have a characteristic wavelength of $0.2$ kpc with a velocity of around $100$km/s which helps to suppress the
formation of small-scale structures. The BEC has solitonic properties thereby forming a solitonic core  at the center of each dark matter halo. According to recent simulations \cite{PhysRevLett.113.261302}, the core radius is comparable to the characteristic wavelength, and the halo transitions to a NFW
profile within a few core radii. Due to the quantum mechanical properties at small scales, it is able to circumvent many of the issues of the usual CDM while preserving the observational constraints at large scales  \cite{2014NatPh..10..496S}. Constraints on FDM from the cosmic microwave background are described in \cite{2015PhRvD..91j3512H}. At the galactic scale, the resultant quantum pressure stabilizes the gravitational collapse, thus preventing the formation of cusp at the centre. In this model, the dark matter halo consists of two components, a BEC core and an outer region with non-condensate DM particles having the well-known NFW (Navarro, Frenk, White) profile. This model has been widely tested in the context of galaxy rotation curves, weak lensing, X-ray emission from colliding clusters \cite{2020PhRvD.101b3508M,Kendall:2019fep}.\\

 The Schrodinger-Poisson Equations describing the dynamics of the system are given by \cite{2017PhRvA..95a3844N}.
\begin{eqnarray}
 i \hbar\frac{\partial \psi}{\partial t}&=&-\frac{\hbar^2}{2m}\nabla^2\psi+mU\psi \\
 \nabla^2U &=&4\pi G\rho
\end{eqnarray}
$U$ is the gravitational potential, $m$ is the mass of the particles.
The total mass density profile of the halo in the FDM model is given by:
 \begin{equation}
 \rho =\rho_{c}\theta(r_t-r)+\rho_{NFW}\theta(r-r_t),
 \label{fuzzy}
 \end{equation}
where $\theta$ is a step function and $r_t$ is  the scale/radius at which the transition from BEC to ordinary CDM phase happens. As in the earlier case, here also the first term corresponds to the core density of the FDM halo profile up to radius $r_t$ whereas the second term denotes the density profile between $r_t$ and the halo radius $R$.
Here
\begin{eqnarray}
    \rho_{c}(r)&=&\rho_{0}\left[1+0.091\left(\frac{r}{r_{c}}\right)^2\right]^{-8}, \label{fuzzycore}\\
    \rho_{0}&=&1.9  \times\left(\frac{10^{-23}\text{eV}}{m}\right)^{2}\left(\frac{\text{kpc}}{r_{c}}\right)^{4} M_{\odot}\text{pc}^{-3},\\
    r_{c}&=&1.6\text{kpc} \times\left(\frac{10^{-23}\text{eV}}{m}\right)(10^3M_{12})^{-1/3}.
\end{eqnarray}
where $\rho_0$ is the central mass density of the core. The NFW profile has been described in the previous section in Eq. \eqref{nfweq}.
Therefore in place of Eq. \eqref{ri}, we have
\begin{eqnarray}
    n=1+ \frac{2\pi G}{w^2R}\left( \rho_0\int_0^{r_t}dr\left[1+0.091\left(\frac{r}{r_{c}}\right)^2\right]^{-8} +\rho_{NFW}^0\int_{r_t}^{R}dr\frac{1}{r/R_{NFW}\Big(1+(r/R_{NFW})^2\Big)}\right) \nonumber \\
\end{eqnarray}
Using these expressions, we now evaluate the relative change in speed $\delta c_{eff}$ similar to the earlier cases.\\

For the calculation, we use the following parameter values as used in \cite{NFW}:\\

$\rho_{NFW}^0=2.9\times 10^{-2}\ M_{\odot}/pc^3; \quad R_{NFW}=10^4 {\rm pc}; \quad R=10^5 {\rm pc};\ r_t=60\times 10^3{\rm pc}$ \\

The distance to the source is taken to be
$D=39.6 \times 10^6 {\rm pc}$, as in the previous case.

\begin{figure}[ht]
\centering
\includegraphics[width=4in]{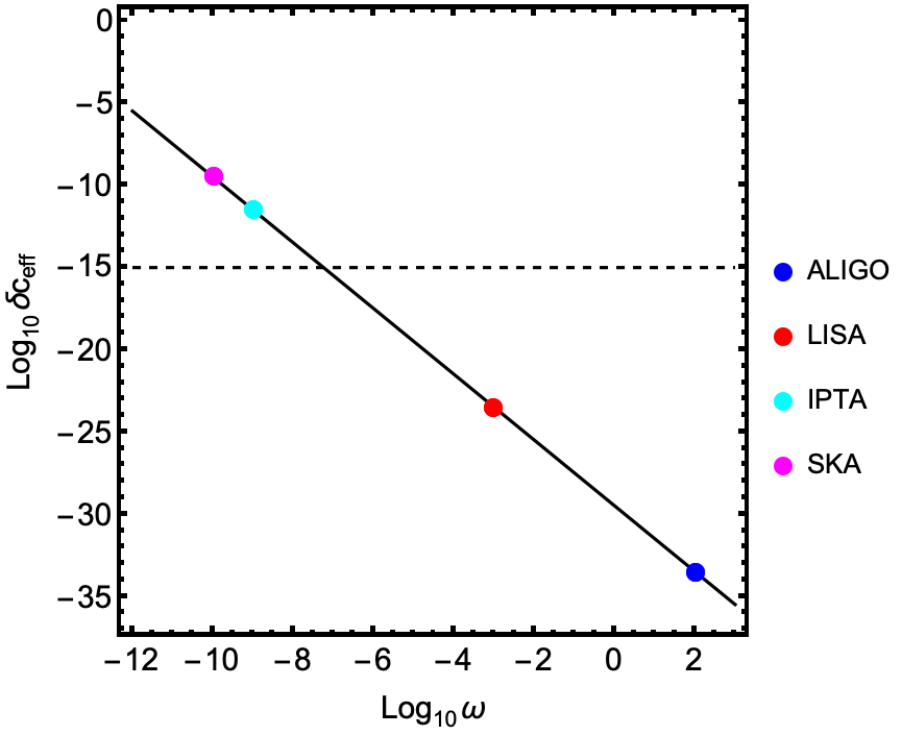}
\caption{{\it Variation of change in the effective speed of GW ($\delta c_{eff}$) with frequency ($\omega$), given by the solid line, for fuzzy dark matter with $m=10^{-22}$ eV. Similar to the superfluid case, we assumed that the ratio of fuzzy core mass to total halo mass is $0.9$. The dotted line corresponds to the current upper limit on $\delta c_{eff}$ put by LVK from the multi-wavelength detection of GW170817 in EM and GW frequencies. The coloured dots represent the change in effective GW speed for ALIGO, LISA, IPTA and SKA detection frequencies.}}
\label{plotfuzzy}
\end{figure}

Here, we have explored the parameter space allowed in the fuzzy dark matter model from the GW speed constraint obtained from the GW170817 data.  The only free model parameter in this model is the axion mass $m$ which typically lies in the range $10^{-21} - 10^{-23}$ eV from various observations like galaxy clusters, large scale structures, galaxy rotation curves, galactic halos, dwarf galaxies \cite{PhysRevD.28.1243,PhysRevLett.64.1084,1994PhRvD..50.3650S,2000PhRvL..85.1158H,Goodman:2000tg,Peebles_2000,2006PhLB..642..192A,2014NatPh..10..496S}. Here we have considered the halo density profile of the fuzzy dark matter as given in \cite{fuzzy1}. Eq. \eqref{fuzzycore} shows the density profile considered in the analysis and the parameter values (for the Milky Way galaxy) used have been listed above. Using this density profile, we evaluate the relative change in the speed of GW and compare it with the constraints from GW170817 similar to the superfluid case. In Fig. \ref{plotfuzzy}, we have shown the effect of detection frequency on the relative change in speed for $m=10^{-22}$ eV. Similar to the superfluid case, here also we considered the fuzzy core mass to be $90\%$ of the total halo mass. The dotted line corresponds to the current upper limit on $\delta c_{eff} \sim 10^{-15}$ put by LVK from the multi-wavelength detection of GW170817, while the coloured dots represent the change in effective GW speed for ALIGO, LISA, IPTA and SKA detection frequencies. The values of $\delta c_{eff}$ for ALIGO, LISA, IPTA and SKA are $\sim 10^{-34}$, $\sim 10^{-24}$, $\sim 10^{-12}$ and $\sim 10^{-10}$ respectively. As can be seen, compared to the superfluid case, both SKA and IPTA are above the detectable range for fuzzy dark matter.

\section{BEC dark matter}
\label{sec5}

The BEC dark matter model is a model of self-interacting DM particles consisting of very light bosonic particles where the Bosons form a condensate whose De-Broglie wavelength is larger than the mean inter-particle distance. The Bose-Einstein condensate (BEC) is a well studied phenomenon in the context of Bosonic systems. Bosons, having an integer spin, can share the same quantum state. If the temperature of a bosonic system is below a certain critical temperature, then the only available energy state for all the particles is the ground state with minimum energy. Under these circumstances, the particles can form a condensate state, called Bose-Einstein Condensate, where the entire system behaves as a macroscopic quantum system. The macroscopic state of the BEC can be described by a single scalar field, which can also be interpreted as a single coherent wavefunction $\psi$ for the macroscopic state. In the context of astrophysics, BEC has been hypothesized to form in high density environments such as the core of a neutron star \cite{Membrado:1989ke,Chavanis:2011cz,Li:2012sg,PhysRevD.91.084051}. In cosmology, it was mainly introduced to overcome the small scale anomalies of the CDM, in particular the core-cusp problem. Since the BEC is a quantum system, at high densities near the center of the DM halo, the quantum pressure arising due to the Heisenberg's uncertainty principle prevents the formation of a cusp, thus mitigating the cuspy-core problem of the normal CDM. The study of BEC and its properties in the astrophysical and cosmological set-up has thus garnered much interest and is currently a very active field of research \cite{PhysRevLett.103.111301,Harko:2011dz,Lee:2008jp,Kain:2010rb,PhysRevD.84.043531,2011PhRvD..84d3532C,10.1111/j.1365-2966.2011.18386.x,Schive:2015kza,10.1093/mnras/stw1256,PhysRevD.95.063515,PhysRevD.101.063532}. The main characteristics of this model are described in terms of two parameters: mass $m$ and the scattering length $a$ of the boson particles. In order for the BEC to be relevant at galactic scales, the de Broglie wavelength has to be very large, or in other words, the mass of the particles has to be very small, typically much less than 1 eV \cite{2012,Marsh:2015xka,Li:2016mmc,Sin:1992bg,Arbey:2001qi,Matos:2000ss}. These are known as ultralight bosons. Such particles have been widely considered to study the properties of DM halo in galaxies and galaxy clusters. It has been proposed that the solitonic core of the BEC DM halo is surrounded by the excited states of the scalar field, described by the phonon modes. 

The effective Lagrangian describing the cosmological dynamics is given as
\begin{equation}
    {\mathcal L}=\frac{1}{2}\partial_{\mu}\phi\partial^{\mu}\phi-\frac12m^2\phi^2-\lambda \phi^4
\end{equation}
Here, $\lambda$ is related to the scattering length $a$ of the particles as,
\begin{equation}
  a = \frac{3\lambda}{2\pi m}  
\end{equation}

One can define a coherent wavefunction for the BEC in terms of the scalar field :
$$ \psi = \phi e^{i mt/\hbar}$$

The Gross-Pitaevskii (GP) equation can be arrived at by expressing the Lagrangian in terms of the wavefunction $\psi$ and taking the non-relativistic limit. Taking the gravitational interaction of the DM particles into account, the resulting equation becomes a system of coupled differential equations, known as the Gross-Pitaevskii-Poisson (GPP) equations. The GPP equations are expressed as follows:

\begin{equation}
    i \hbar\frac{\partial \psi}{\partial t}=-\frac{\hbar^2}{2m}\nabla^2\psi+m\Phi\psi + \frac{4\pi a \hbar^2}{m^2}|\psi|^2\psi
\end{equation}
\begin{equation}
    \nabla^2\Phi = 4\pi G |\psi|^2
\end{equation}

where $\Phi$ is the Newtonian gravitational potential.\\
From the GPP equation, the mass density $\rho$ can be defined as $\rho = |\psi|^2 $. An equivalent description can be given in terms of hydrodynamic equations by defining the various quantities in the language of fluid dynamics. This can be achieved by performing a variable transformation, known as the Madelung transformation. For maintaining brevity of the paper, we refrain from outlining the detailed derivation of the fluid equations here. Interested readers may refer to the various literature discussing the derivation \cite{Chavanis:2018pkx,Sreenath:2018ple,Boehmer:2007um}. The resulting fluid equations can be compared with the continuity and Euler equations of a fluid with density $\rho$, velocity $\vec{u}$ and pressure $P$.
With this direct comparison, the pressure and density are found to be related to each other by the following relation
\begin{equation}
    P=\frac{2\pi a \hbar^2}{m^3}\rho^2
    %\label{eos}
\end{equation}
which represents a two-body interaction i.e. the properties of the system can be described by a potential involving the separation between any two particles. In the case of BEC, the potential arising due to the self-interaction of the particles can be represented as a sum of pair-wise interaction terms.

This gives rise to a Lane–Emden type nonlinear second order differential equation for the density profile ($n=1$) whose exact analytical solution can be obtained which gives the density profile for this model. The corresponding form for self interacting dark matter halo forming non rotating BEC with spherical symmetry is given by
\begin{equation}
    \rho(r)=\rho_0\frac{\sin kr}{kr}
    \label{bec}
\end{equation}
Here, $\rho_0$ is the integration constant which is set equal to the central density $\rho_c$, and $k$ is a constant involving the length scale of the theory, $k= \sqrt{\frac{Gm^3}{a\hbar^2}}$.\\
Note that, the solution is non-singular at $r=0$, and thus evades the problem of a cuspy core. 

As discussed for the other two cases, the total DM halo consists of two parts: a BEC core and a CDM crust. The halo density profile is again given by \eqref{fuzzy}.
The core radius $r_{c}$ can be obtained by using the boundary condition $\rho_c(r_c)=0$. Similarly, using the condition $\int \rho_c d^{3}r = M_{\rm core}$, $M_{\rm core}$ being the BEC core mass, one gets the central density $\rho_0$.
The corresponding expressions for the radius of the core and central density $\rho_0$ in terms of the mass $m$ of the BEC particle and scattering length $a$ are as follow \cite{bec1}:
\begin{eqnarray}
    r_c&=&\frac{\pi}k=\pi \sqrt{\frac{a\hbar ^2}{Gm^3}}\nonumber\\
&&=13.5\times \left(\frac{a}{10^{-17}\;{\rm cm}}\right)^{1/2}\times \left(\frac{m}{10^{-36}\;{\rm g}}\right)^{-3/2}\times 10^3\;{\rm pc} \nonumber\\
&&=5.512\times10^{7} \times \left(\frac{a}{{\rm cm}}\right)^{1/2}\times \left(\frac{m}{{\rm eV}}\right)^{-3/2} \;{\rm pc} \nonumber \\
\label{radiusbec}\\
\rho_0&=&\frac{\pi M_{\rm core}}{4 r_c^3}\;{\rm \mathrm{M_\odot}/pc^{3}}
\label{centralbec}
\end{eqnarray}
Thus for this case, $a,\ m$ are the two model parameters which are to be constrained from observations.

Substituting the form \eqref{bec} in Eq. \eqref{5b} and using \eqref{fuzzyold}, we obtain the refractive index as

\begin{eqnarray}
    n=1+\frac{2\pi G}{w^2R}\left(\rho_0\int_0^{r_t}dr\frac{\sin kr}{kr}+\rho_{NFW}^0\int_{r_t}^{R}dr\frac{1}{r/R_{NFW}\Big(1+(r/R_{NFW})^2\Big)}\right)
\end{eqnarray}

The above integration is performed numerically and the results are then fed into Eq. \eqref{ceff} to obtain the constraint on the model parameters similar to the earlier cases.

\begin{figure}[h!]
\centering
\includegraphics[width=4in]{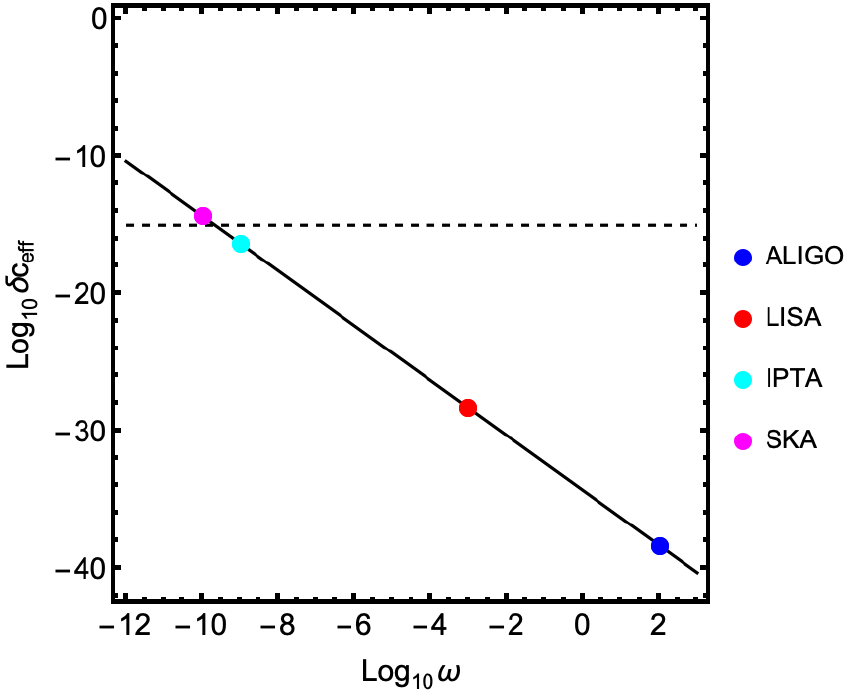}
\caption{{\it Variation of change in the effective speed of GW ($\delta c_{eff}$) with frequency ($\omega$), given by the solid line, for BEC dark matter with $a=10^{-20}$ cm and $m=10^{-5}$ eV. The ratio of the BEC core mass to total halo mass is $0.9$. The dotted line corresponds to the current upper limit on $\delta c_{eff} \sim 10^{-15}$. The coloured dots represent the change in effective GW speed for ALIGO, LISA, IPTA and SKA detection frequencies.}}
\label{plotbec1}
\end{figure}

\begin{figure}[h!]
\centering
\includegraphics[width=5in]{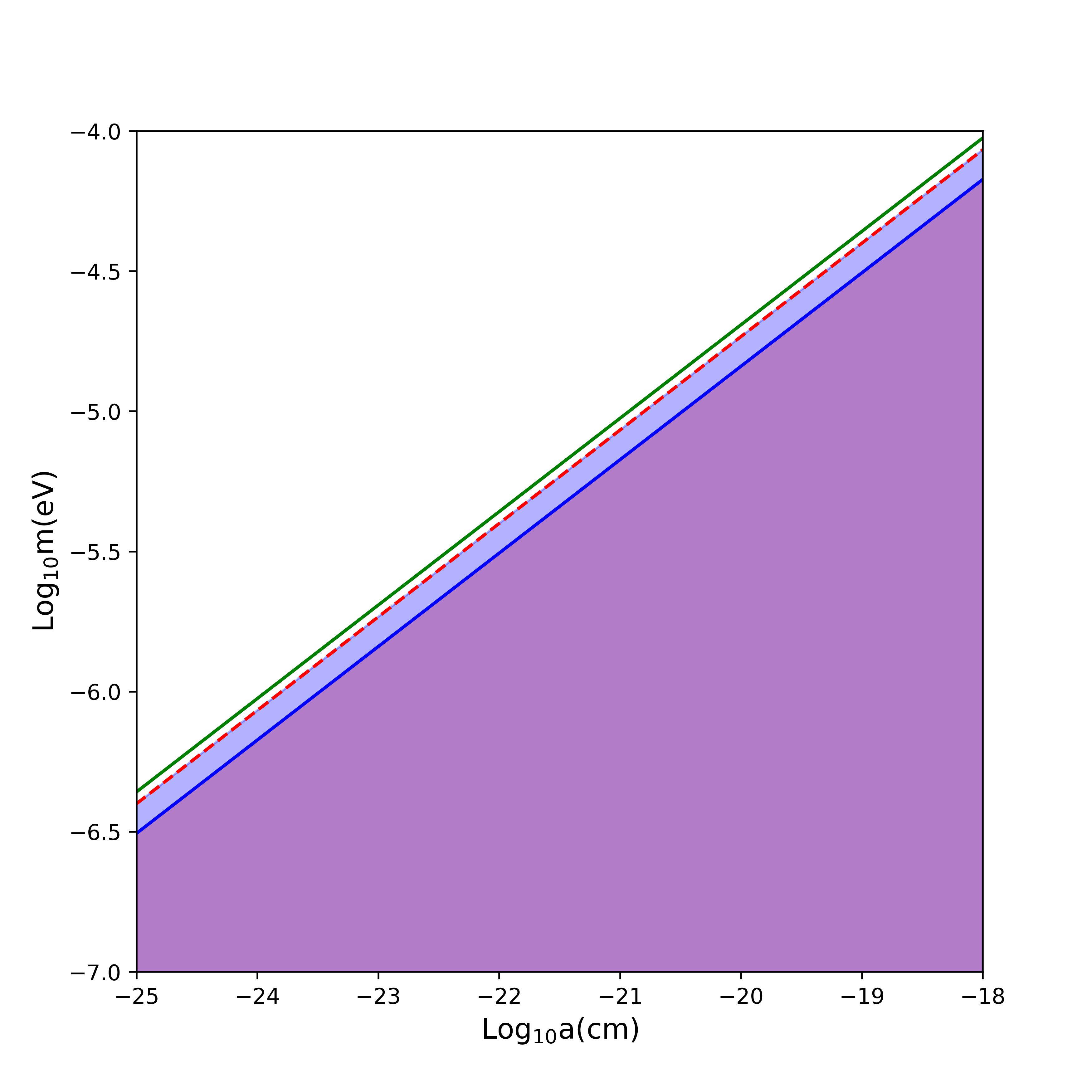}
\caption{{\it Allowed parameter space for BEC dark matter with frequency fixed in the nanohertz range. The blue line represents the parameter values defined by $R=100$kpc and the green line is for $R=60$kpc. The red dashed line corresponds to the upper limit imposed by the GW speed constraint. The shaded area below this line represents the region allowed by the current constraint on the relative change in speed. The blue shaded region in between the red and the blue lines represent the region allowed by both speed constraint and radius constraint. }}
\label{plotbec}
\end{figure}

In the case of BEC, we use the density profile given in Eq. \eqref{bec}, which has also been discussed in \cite{bec1}. This model has two free parameters: $m$, the axion mass and $a$, the scattering length. Repeating the same analysis as in the earlier cases, in Fig. \ref{plotbec1}, we have shown the effect of detection frequency on $\delta c_{eff}$ for parameter values $m=10^{-5}$eV and $a=10^{-20}$ cm. Similar to the earlier cases, we took the BEC core mass to be $90\%$ of the total halo mass. The dotted line denotes the upper limit on $\delta c_{eff}$ put by LVK from the multi-wavelength detection of GW170817. The coloured dots represent the change in effective GW speed for ALIGO, LISA, IPTA and SKA detection frequencies. From the plot one can see that the behaviour is very close to what we saw for the superfluid case. As we can see, the values of $\delta c_{eff}$ for ALIGO and LISA are far below the upper cut-off for $\delta c_{eff}$. However,  $\delta c_{eff}$ for IPTA and SKA detectable frequencies are in and around the current threshold of the measured GW speed, hence they can be used to constrain the relevant model parameters.

 In Fig. \ref{plotbec}, we further plot the allowed parameter space for the BEC dark matter considering the best case scenario where the frequency is fixed in the nanoHz range. We show the parameter space allowed out by the relative change in speed. Similar to the superfluid case,  we assume the transition radius to be at $r_t \sim 60$ kpc \cite{Berezhiani:2017tth}.  
 From \eqref{radiusbec}, for a given radius, we get a relation between $a$ and $m$. Once again, requiring that the core radius has to be less than the Milky Way radius ($100$ kpc.), but greater than the transition radius $60$ kpc, we get additional constraints on the model parameters. The blue line indicates the parameter values defined by $R=100$ kpc and the green line corresponds to $R=60$ kpc. The red dashed line corresponds to the upper limit on $\delta c_{eff}\sim 10^{-15}$ imposed by the GW speed constraint. The shaded region below this line indicates the parameter space allowed by the speed constraint alone. The blue shaded region in between the red and blue lines represents the area allowed by both speed and radius constraints. Thus, we find, similar to the superfluid case, there is a small region containing parameter pairs that are allowed by the radius constraint but discarded from the relative speed constraint. Thus, the speed of the gravitational wave puts additional constraints on the model parameters.
 
\begin{figure}[h!]
\centering
\includegraphics[width=4in]{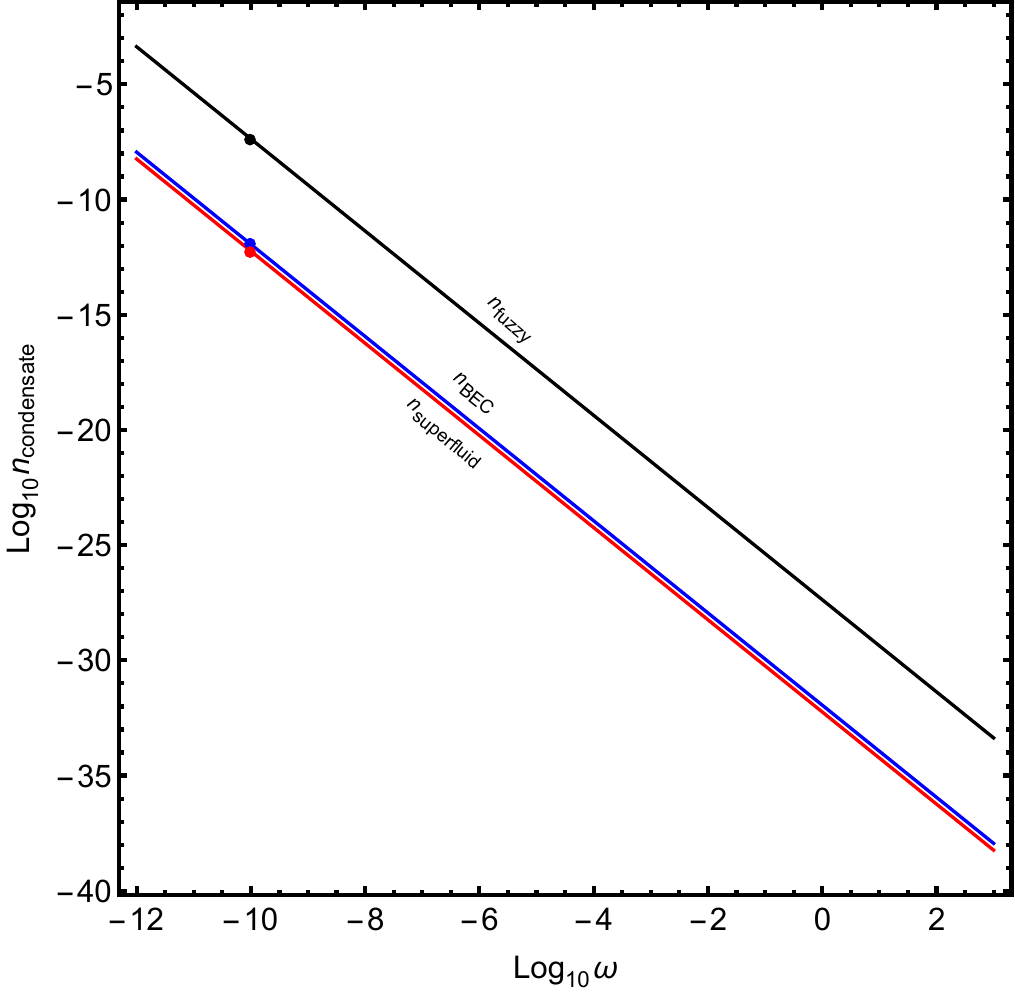}
\caption{{\it Variation of change in the refractive index of the condensate ($n_{condensate}$ for the three models (superfluid, BEC, fuzzy dark matter) with frequency ($\omega$). The black line is for fuzzy dark matter for $m=10^{-22}$eV, the red line is for BEC dark matter with parameter values $a=10^{-20}$cm, $m=10^{-5}$eV and the blue line corresponds to superfluid dark matter with/without baryonic interaction for $\Lambda=10^3$meV, $m=5\times 10^{-2}$eV. The black, red and blue dots represent the values of the refractive index at $\omega=10^{-10}$Hz for fuzzy, BEC and superfluid dark matter respectively.}}
\label{n-w}
\end{figure}

\section{Discussions}
\label{sec6}

In this work, we have analysed three different dark matter models, namely, the superfluid dark matter, the BEC dark matter and the fuzzy dark matter in the light of gravitational wave propagation through them. In particular, we have used the GW speed constraint as obtained from the BNS GW170817 by the LIGO-Virgo detector network and used it to study the allowed parameter space for these three models for ALIGO, LISA, IPTA and SKA detection frequencies. In all the above cases, we make use of the fact that the GW travels with a reduced speed through these media, in comparison to the speed of light in vacuum. Using the current constraint on the relative change in speed of the GW for a nearby source like GW170817, we find that the effect of a superfluid core in the galactic halo is extremely small for Advanced LIGO-Virgo detection frequency range, even for the most optimistic case where the superfluid core mass is assumed to contain $90\%$ of the total halo mass. We note that the effect on the GW speed is most sensitive to the change in detection frequency. Thus, going to a lower frequency magnifies the overall effect by enhancing the relative change in speed. Of particular interest is detectability at frequencies in the nanoHz range, for which the relative GW speed exceeds the current threshold of $10^{-15}$. We show this dependence on frequency in Fig. \ref{plot1} and show where the current/proposed detector sensitivities lie. The most promising candidates appear to be the radio telescopes such as PTA and SKA that will be able to detect gravitational waves at much lower frequencies. Unless a more precise measurement of the relative speed of GW is possible with the planned detectors such as Einstein Telescope \cite{Maggiore:2019uih}, Cosmic Explorer \cite{Reitze:2019iox} or LISA, the current generation of ground based detector network would not be able to resolve the presence/ absence of a superfluid core at the center of the DM halo. We also investigate the allowed parameter space for this model for a given frequency $\sim 10^{-9}$ Hz. The parameter space is already narrowed down by the radius bound i.e. the radius of the core has to be less than the total halo radius ($100$ kpc) and larger than the transition radius ($60$ kpc), which defines the transition from superfluid to NFW profile (see section \ref{sec3} for discussion on this). The GW speed constraint is shown using shaded region. The blue shaded region is allowed by both radius and speed constraint. Thus, the GW speed effect does provide new constraints albeit very narrow when the bounds coming from the maximum allowed radius is considered. We also analysed the effects mentioned above for the case when the superfluid-baryon interaction is present, by solving the modified Lane-Emden equation for different values of  the interaction strength $\alpha$ within its current bound \cite{Berezhiani:2015bqa}. In this case, we find that the baryonic interaction has no effect on the overall density profile, and hence on the GW speed. This is because the propagation speed of GW is not affected by the nature of interaction between the two components, rather it depends on the density distribution of the medium which is dominated by the first term in Eq. \eqref{int}. Mathematically this can be understood since the central core density has no direct dependence on the interaction strength $\alpha$ in Eq. \eqref{4}.

For the case of fuzzy dark matter, the mass range that we consider is already constrained by other observations, as cited in section \ref{sec4}. Since the allowed masses are already very low ($\sim 10^{-22}$ eV), the effect on GW speed seems to be much more significant compared to the superfluid model. A similar $c_{eff}-\omega$ plot in this case is shown in Fig. \ref{plotfuzzy} for a fixed mass $m=10^{-22}$ eV. As we can expect, the change in GW speed is well above the threshold for both IPTA and SKA detection frequencies, much higher than that in the superfluid scenario. This implies that this model can be falsified/validated at a higher detection frequency as compared to the superfluid model, and can thus be tested more easily.

For the BEC model, there are two free model parameters $m$ and $a$ as discussed in section \ref{sec5}. We study the effect of BEC core and an NFW outer crust on the GW speed in the same way as for the other two cases discussed above. For a representative value of the model parameters ($m=10^{-5}$ eV and $a = 10^{-20}$ cm), the dependence of the speed on the detection frequency is shown in Fig. \ref{plotbec1}. The effect of BEC on the GW speed seems to be similar to superfluid core but weaker than the fuzzy DM scenario. For a fixed frequency around nanoHz, we also study the parameter space of BEC along with the two radius bounds coming from the overall halo radius and the transition radius in a similar fashion as described above for the superfluid case. This is shown in Fig. \ref{plotbec} where the two solid lines correspond to the bounds allowed by the radii observation. Within the narrow region allowed by this constraint, we get additional constraint obtained from the effect of BEC on relative change in speed. The blue shaded region corresponds to the parameter values allowed by both speed and radius constraints.

For all the above cases we have assumed the core mass to be $90\%$ of the total mass. Since the Fuzzy dark matter core exhibits the strongest effect for a given frequency, we use this model to investigate the effect of other parameters such as the core mass fraction and the distance to the source on the GW speed. We studied different core mass fractions ranging between $10\% - 90\%$ of the total mass and found that this variation contributes to roughly $10\%$ change in the GW speed, which is almost linear in the logarithmic scale. Additionally, the dependence of $\delta c_{eff}$ on the source distance have been found to be approximately inversely proportional in the logarithmic scale i.e. if the source distance is increased by 10 orders of magnitude, the change in the relative GW speed reduces roughly by a factor of 10. So, this effect is not much of concern unless we are dealing with very high redshift observations within the LIGO-Virgo frequency band. In other words, the dependence on these factors are much weaker compared to the frequency dependence. For a superfluid core, we found that our analysis is nearly independent of these two parameters i.e. the results do not vary much if we change the distance to the source or change the relative mass of the superfluid core.

We also compared the contribution of the core vs. the NFW crust to the change in refractive index for each of the cases considered. The ratio of the contribution from core to that from the ordinary DM ($n_{condensate}/n_{NFW}$) is the weakest for the superfluid and BEC scenarios and turn out to be of the order of 10. It is found to be the strongest for fuzzy DM for which the value is $\sim 10^6$.
In Fig. \ref{n-w}, we have further compared how the refractive index for the core varies with the frequency for the three models. As expected, fuzzy dark matter has the highest contribution to the refractive index. The contributions of superlfuid and BEC are nearly the same and is much weaker as compared to fuzzy dark matter. The three dots shown in this figure correspond to a detection frequency $\omega = 10^{-10}$ Hz and shows the contribution of the core to the refractive index for each of the three models at this frequency. This value of the frequency is chosen because this is best case scenario for which these models can be tested. From this plot, one can conclude that out of the three condensate scenarios, the fuzzy DM model is the most feasible scenario to be tested in near future with GW detectors. With the data releases and analyses that IPTA has already begun \cite{Antoniadis:2022pcn}, testing the viability of these models does not seem too far-fetched. 

\section*{Acknowledgements}
S. Bera would like to acknowledge support from the Universitat de les Illes Balears; European Union FEDER funds; the Spanish Ministry of Science and Innovation and the Spanish Agencia Estatal de Investigación grants PID2019-106416GB-I00/   MCIN/   AEI/10.13039/501100011033, RED2018-102661-T, RED2018-102573-E, FPA2017-90566-REDC; the European Union NextGenerationEU (PRTR-C17.I1); the Comunitat Autònoma de les Illes Balears through the Direcció General de Política Universitària i Recerca with funds from the Tourist Stay Tax Law ITS 2017-006 (PRD2018/24, PDR2020/11); the Conselleria de Fons Europeus, Universitat i Cultura del Govern de les Illes Balears; the Generalitat Valenciana (PROMETEO/2019/071); and EU COST Actions CA18108, CA17137, CA16214, and CA16104. S. Bera would also like to thank IUCAA for providing resources where part of this work was carried out. DFM thanks the Research Council of Norway for their support.  Computations were performed on resources provided by UNINETT Sigma2 -- the National Infrastructure for High Performance Computing and Data Storage in Norway. S. Banerjee would like to thank Friedrich Alexander University for post doctoral research support during the completion of this work.

\bibliographystyle{JHEP} 
\bibliography{main}

\end{document}